\documentclass{iopart}
\usepackage{euscript,iopams,amssymb,amsfonts,graphicx,bm}
\usepackage{pgfplots,setstack}

\usepackage{float}
\usepackage{epsfig}
\usepackage{esint}
 \usepackage{tikz-cd}

\usepackage{bm,braket}

\eqnobysec

\newcommand{\dis}{\displaystyle}

\newcommand{\calU}{{\mathcal U}}
\newcommand{\calG}{{\mathcal G}}

\newcommand{\calH}{{\mathcal H}}

\newcommand{\calP}{{\mathcal P}}
\newcommand{\calQ}{{\mathcal Q}}
\newcommand{\calA}{{\mathcal A}}
\newcommand{\calM}{{\mathcal M}}
\newcommand{\calF}{{\mathcal F}}
\newcommand{\R}{{\mathbb R}}

\newcommand{\X}{\mathbf{X}}
\newcommand{\ellb}{\bm \ell}
\renewcommand{\k}{{\bf k}}

\renewcommand{\P}{\mathbb{P}}
\newcommand{\PP}{\widetilde{\calP}}
\newcommand{\QQ}{\widetilde{\calQ}}
\newcommand{\x}{\mathbf{x}}
\newcommand{\y}{\mathbf{y}}
\renewcommand{\e}{{\mathrm e}}
\newcommand{\E}{{\mathbb E}}

\newcommand{\n}{\mathbf n}
\newcommand{\calT}{{\mathcal T}}
\renewcommand{\P}{\mathbb P}

\newcommand{\q}{\mathbf{q}}

\newcommand{\G}{\widetilde{G}}
\renewcommand{\H}{\widetilde{H}}
\newcommand{\J}{\widetilde{J}}

\begin{document}


\title[Encounter-based reaction-subdiffusion model  II]{Encounter-based reaction-subdiffusion model II: partially absorbing traps and the occupation time propagator}

\author{Paul C. Bressloff}
\address{Department of Mathematics, University of Utah 155 South 1400 East, Salt Lake City, UT 84112}

\begin{abstract} 
In this paper we develop an encounter-based model of reaction-subdiffusion in a domain $\Omega$ with a partially absorbing interior trap $\calU\subset \Omega$. We assume that the particle can freely enter and exit $\calU$, but is only absorbed within $\calU$. We take the probability of absorption to depend on the amount of time a particle spends within the trap, which is specified by a Brownian functional known as the occupation time $A(t)$.  The first passage time (FPT) for absorption is identified with the point at which the occupation time crosses a random threshold $\widehat{A}$ with probability density $\psi(a)$. Non-Markovian models of absorption can then be incorporated by taking $\psi(a)$ to be non-exponential. The marginal probability density for particle position $\X(t)$ prior to absorption depends on $\psi$ and the joint probability density for   the pair $(\X(t),A(t))$, also known as the occupation time propagator. 
In the case of normal diffusion one can use a Feynman-Kac formula to derive an evolution equation for the  propagator. However, care must be taken when combining fractional diffusion with chemical reactions in the same medium. Therefore, we derive the occupation time propagator equation from first principles by taking the continuum limit of a heavy-tailed CTRW. We then use the solution of the propagator equation to investigate conditions under which the mean FPT (MFPT) for absorption within a trap is finite. We show that this depends on the choice of threshold density $\psi(a)$ and the subdiffusivity. Hence, as previously found for evanescent reaction-subdiffusion models, the processes of subdiffusion and absorption are intermingled.

\end{abstract}

\maketitle

\newpage

\section{Introduction}

This is the second of a pair of papers concerned with encounter-based reaction-subdiffusion models. In our first paper \cite{PCBI}, we focused on the case of subdiffusion in a bounded domain $\Omega$ whose surface $\partial \Omega $ was partially absorbing. Following previous encounter-based models for normal diffusion \cite{Grebenkov20,Grebenkov22,Bressloff22,Bressloff22b,Grebenkov22a}, we assumed that the probability of adsorption depended upon the amount of particle-surface contact time; the latter was determined by a Brownian functional known as the boundary local time $\ell(t)$ \cite{Ito63,McKean75,Majumdar05}. We equated the first passage time (FPT) for adsorption with the point at which the local time crossed a randomly generated threshold $\widehat{\ell}$. Different models of adsorption (Markovian and non-Markovian) then corresponded to different choices for the random threshold probability density $\psi(\ell)$. We showed how the marginal probability density for particle position prior to adsorption could be determined in terms of $\psi$ and the joint probability density for particle position $\X(t)$ and the local time $\ell(t)$, also referred to as the local time propagator. We derived an evolution equation for the local time propagator by taking the continuum limit of an analogous propagator equation for a continuous-time random walk (CTRW) with a heavy-tailed waiting time density. Laplace transforming the local time propagator equation with respect to $\ell$ resulted in a fractional diffusion equation, which was supplemented by a Robin or radiation boundary condition on $\partial \Omega$ whose reactivity constant was the Laplace variable $z$ conjugate to the local time. Finding the inverse Laplace transform of the solution with respect to $z$ then yielded the local time propagator. We used our model to investigate the effects of subdiffusion and non-Markovian adsorption on the long-time behavior of the FPT density, extending previous studies of subdiffusion with Dirichlet or Robin boundary conditions \cite{Condamin07,Yuste07,Grebenkov10}.

In this paper we turn to the complementary problem of reaction-subdiffusion in a domain $\Omega$ with a partially absorbing interior trap $\calU\subset \Omega$. If the particle cannot enter the trap but is absorbed on the trap surface $\partial \calU$, see Fig. \ref{fig1}(a), then we recover the type of problem considered in Ref. \cite{PCBI}. Here, however, we assume that the particle can freely enter and exit $\calU$, but is only absorbed within $\calU$, see Fig. \ref{fig1}(b). The main difference from encounter-based models of surface adsorption is that the relevant Brownian functional is now the occupation time $A(t)$, which tracks the amount of time the particle spends within $\calU$. Furthermore, the FPT for absorption is identified with the point at which the occupation time crosses a random threshold $\widehat{A}$.  In the case of normal diffusion one can use a Feynman-Kac formula to derive an evolution equation for the corresponding occupation time propagator, which is the joint probability density for $\X(t)$ and $A(t)$ \cite{Bressloff22,Bressloff22a,Bressloff22c}. Laplace transforming the occupation time propagator equation leads to a diffusion equation with a constant rate of absorption $z$ in $\calU$ and no absorption in the complementary domain. Since $z$ is the Laplace variable conjugate to the occupation time, the inverse Laplace transform yields the occupation time propagator. This can be combined with the probability density $\psi(a)$ of the occupation time threshold $\widehat{A}$ to determine the marginal density for particle position prior to absorption.

Developing an analogous encounter-based model of a reaction-subdiffusion process with a partially absorbing trap is non-trivial. This is a consequence of the fact that considerable care has to be taken in combining anomalous diffusion with chemical reactions occurring within the same complex medium \cite{Henry00,Henry06,Sokolov06,Yuste07,Henry08,Abad10,Yuste10,Fedotov10,Henry13}. In particular, one cannot simply write down an evolution equation with separate diffusion and reaction terms, in which the diffusion term is exactly the same with or without the chemical reaction. Indeed, failure to properly account for the effects of chemical reactions on the fractional diffusion operator may yield unphysical results, including negative particle concentrations. Therefore, we derive the occupation time propagator equation from first principles by taking the continuum limit of a heavy-tailed CTRW, following along analogous lines to previous studies of fractional diffusion equations with first-order death processes or evanescence \cite{Yuste07,Abad10,Yuste10}. We then use the solution of the propagator equation to investigate conditions under which the mean FPT (MFPT) for absorption within a trap is finite. We show that this depends on the choice of threshold density $\psi(a)$ and the subdiffusivity.
 
The structure of the paper is as follows. In section 2 we briefly review the encounter-based model of  reaction-diffusion in the presence of a partially absorbing trap, as developed elsewhere \cite{Bressloff22,Bressloff22a}.
In section 3 we derive the fractional diffusion equation for the occupation time propagator in the case of reaction-subdiffusion. We first construct a Feynman-Kac equation for the propagator of a corresponding  CTRW. The fractional diffusion equation is then obtained by taking the continuum limit in the case of a heavy-tailed waiting time density. We then show how Laplace transforming the propagator equation with respect to the occupation time generates a reaction-subdiffusion equation that is identical in form to an evanescent fractional diffusion equation \cite{Yuste07,Yuste10,Abad10}. Given a solution of the latter, we can incorporate a non-Markovian model of absorption by inverting the Laplace transform and introducing the density $\psi(a)$. Finally, in section 4 we apply our theory to the case of a single spherical trap. That is, we represent $\calU$ and $\Omega$ as concentric $d$-dimensional spheres with normal diffusion in $ \Omega\backslash \calU$ and subdiffusion within the trap. We calculate the MFPT for absorption and use this to determine how the MFPT (if it exists) depends on $\psi$ and subdiffusion. In particular, we find that subdiffusion and absorption are intermingled. 

  \section{Encounter-based model of a partially absorbing trap: normal diffusion}
  
  \begin{figure}[t!]
\raggedleft
\includegraphics[width=12cm]{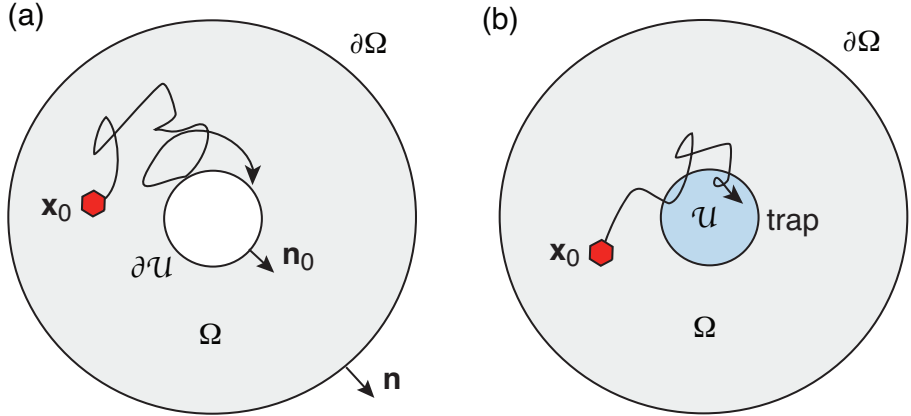} 
\caption{Diffusion in a bounded domain $\Omega$ with a partially absorbing interior trap $\calU$. (a) Adsorption occurs at the surface $\partial \calU$ without the particle ever entering the interior $\calU$. (b) Absorption occurs within $\calU$ and prior to absorption, the particle can freely enter and exit the trap.}
\label{fig1}
\end{figure}

  In order to develop an encounter-based model of reaction-subdiffusion in the presence of a partially absorbing trap, we begin by briefly reviewing the case of normal diffusion \cite{Bressloff22,Bressloff22a}.
  Consider a particle diffusing in a bounded domain $\Omega\subset \R^d$ with a totally reflecting boundary $\partial \Omega$ and a partially absorbing interior trap as shown in Fig. \ref{fig1}(b). For the moment, suppose that the particle can freely enter and exit $\calU$ without being adsorbed. For the sake of generality, we allow the diffusivity to be different in the interior and exterior of $\calU$.
 Let $\X(t)$ denote the position of the particle at time $t$. The amount of time the particle spends within $\calU$ over the time interval $[0,t]$ is specified by a Brownian functional known as the occupation time $A(t)$:
\begin{equation}
\label{occ}
A(t)=\int_{0}^tI_{\calU}(\X(\tau))d\tau= \int_{0}^t\int_{\calU}\delta(\X(\tau)-\y)d\y d\tau.
\end{equation}
Here $I_{\calU}(\x)$ denotes the indicator function of the set $\calU\subset \Omega$, that is, $I_{\calU}(\x)=1$ if $\x\in \calU$ and is zero otherwise. Let $P(\x,a,t)$ denote the joint probability density of the pair $(\X(t),A(t))$ at time $t$ for all $\x\in \Omega \backslash \calU$ and let $Q(\x,a,t)$ denote the corresponding density for all $\x \in \cal U$. We refer to the pair $(P,Q)$ as the occupation time propagator of the diffusion process. Using a Feynman-Kac formula, we can derive the following evolution equations for the propagator:
\numparts
\begin{eqnarray}
\label{Pocca}
\fl &\frac{\partial P(\x,a,t)}{\partial t}=D\nabla^2 P(\x,a,t), \ \x \in \Omega\backslash \calU,\\
\label{Poccb}
 \fl &\nabla P(\x,a,t) \cdot \n=0, \  \x \in \partial \Omega, \\
 \label{Poccc}
\fl & \frac{\partial Q(\x,a,t)}{\partial t}=\overline{D}\nabla^2 Q(\x,a,t) -\left (\frac{\partial Q}{\partial a}(\x,a,t) +\delta(a)Q(\x,0,t) \right ),\quad \x \in \calU,
\end{eqnarray}
with $\n$ the unit outward normal on $\partial \Omega$, $D$ the diffusivity in $\Omega\backslash \calU$, and $\overline{D}$ the diffusivity in $\calU$.
We also have the continuity conditions
	\begin{equation}
	\label{Poccd}
\fl 	P(\x,a,t)=Q(\x,a,t),\quad D\nabla P(\x,a,t)\cdot \n_0 =\overline{D}\nabla Q(\x,a,t) \cdot \n_0 ,\   \x \in \partial \calU,
	\end{equation}
	\endnumparts
	and the initial conditions $P(\x,a,0)=\delta(\x-\x_0)\delta(a)$, $Q(\x,a,0)=0$. The unit normal $\n_0$ on $\partial \calU$ is directed towards the exterior of $\calU$, see Fig. \ref{fig1}. We assume that the particle starts out in the non-absorbing region. (The analysis is easily modified if $\x_0\in \calU$.)
	
We now introduce a probabilistic model of partial absorption within $\calU$ by introducing the stopping time 
\begin{equation}
\label{TA}
{\mathcal T}=\inf\{t>0:\ A(t)>\widehat{A}\},
\end{equation}
where $\widehat{A}$ is a random occupation time threshold with probability distribution $\P[\widehat{A}>a]=\Psi(a)$. We identify ${\mathcal T}$, which is the time at which $A(t)$ first crosses the threshold $\widehat{A}$, as the FPT for absorption. The marginal probability density for particle position $\X(t) \in \Omega\backslash \calU$ is then
\[p^{\Psi}(\x,t)d\x=\P[\X(t) \in (\x,\x+d\x), \ t < {\mathcal T}].\]
Since $\calA(t)$ is a nondecreasing function of time, the condition $t < {\mathcal T}$ is equivalent to the condition $A(t)<\widehat{A}$, that is
$p^{\Psi}(\x,t)d\x=\P[\X(t)\in (\x,\x+d\x), \ \calA(t)<\widehat{A}]$. Hence,
\begin{eqnarray}
\fl p^{\Psi}(\x,t)&=\int_0^{\infty}da \psi(a) \int_0^a da'P(\x,a',t)=\int_0^{\infty}da' P(\x,a',t)\int_{a'}^{\infty}da\psi(a) ,\nonumber\\
\fl &=\int_0^{\infty}\Psi(a)P(\x,a,t)da,\quad \x \in \Omega \backslash \calU,
\label{peep}
\end{eqnarray}
where we have reversed the order of integration and set $\psi(a)=-d\Psi(a)/da$.
Similarly,  
\begin{eqnarray}
\label{qeep}
q^{\Psi}(\x,t|\x_0)&=\int_0^{\infty}\Psi(a) Q(\x,a,t)da,\quad \x \in \calU .
\end{eqnarray}
First suppose that $\widehat{A}$ is exponentially distributed so that $\Psi(a)=\e^{-z a}$ for constant $z$. Equations (\ref{peep}) and (\ref{qeep}) imply that $p(\x,t)$ and $q(\x,t)$ (after dropping the superscript $\Psi$) are Laplace transforms of the propagator with respect to $a$:
\numparts
\begin{eqnarray}
 p(\x,t)&=\int_0^{\infty}\e^{-za}P(\x,a,t)da:=\calP(x,z,t),\\ q(\x,t)&=\int_0^{\infty}\e^{-za}Q(\x,a,t)da :=\calQ(\x,z,t).
\end{eqnarray}
\endnumparts
The generators $\calP$ and $\calQ$ satisfy the equations
\numparts
\begin{eqnarray}
\label{za}
\fl &\frac{\partial \calP(\x,z,t)}{\partial t}=D\nabla^2 \calP(\x,z,t),\quad \x \in \Omega \backslash \calU,\\
\label{zb}
\fl &\nabla \calP(\x,z,t) \cdot \n=0, \  \x \in \partial \Omega, \\
\label{zc}
\fl & \frac{\partial \calQ(\x,z,t)}{\partial t}=\overline{D}\nabla^2 \calQ(\x,z,t)-z\calQ(\x,z,t),\quad \x \in \calU  , \\
 \label{zd}
\fl &\calP(\x,z,t)=\calQ(\x,z,t),\quad D\nabla \calP(\x,z,t)\cdot \n_0 =\overline{D}\nabla \calQ(\x,z,t) \cdot \n_0 ,\   \x \in \partial \calU.
\end{eqnarray}
\endnumparts 
Equations (\ref{za})--(\ref{zd}) describe diffusion in a domain containing a trap with a constant absorption rate $z$, and can be solved using standard methods. 
Finally, given $\calP(\x,z,t)$ and $\calQ(\x,z,t)$, the densities $p^{\Psi}(\x,t)$ and $q^{\Psi}(\x,t)$ for non-exponential $\Psi$ can be obtained by inverting the solutions with respect to $z$:
\begin{eqnarray}
     \label{zint}
 \fl  p^{\Psi} (\x,t)&=\int_0^{\infty} \Psi(a){\mathcal L}_{a}^{-1}[\calP(\x,z,t)]da ,\ q^{\Psi} (\x,t)&=\int_0^{\infty} \Psi(a){\mathcal L}_{a}^{-1}[\calQ(\x,z,t)]da,
  \end{eqnarray}
  where ${\mathcal L}^{-1}$ denotes the inverse Laplace transform. 

One general quantity of interest is the survival probability $S^{\Psi}(t)$ that the particle hasn't been absorbed up to time $t$, given that it started at $\x_0$. In the case of a partially absorbing trap $\calU$,
\begin{equation}
\label{Socc}
S^{\Psi}(t)= \int_{\Omega\backslash \calU} p^{\Psi}(\x,t)d\x+\int_{ \calU}  q^{\Psi}(\x,t)d\x.
\end{equation}
Differentiating both sides with respect to $t$ and using equations (\ref{Pocca})--(\ref{Poccd}) gives
\begin{eqnarray}
\fl \frac{dS^{\Psi}(t)}{dt}&= \int_{\Omega\backslash \calU} \frac{\partial p^{\Psi}(\x,t)}{\partial t}d\x+\int_{ \calU}  \frac{\partial q^{\Psi}(\x,t)}{\partial t}d\x\nonumber \\
\fl &=\int_0^{\infty}\Psi(a)\bigg \{\int_{\Omega\backslash \calU}  D\nabla^2 P(\x,a,t)d\x+
\int_{ \calU} \overline{D}\nabla^2 Q(\x,a,t)d\x\nonumber \\
\fl &\quad  -\int_{ \calU}\left (\frac{\partial Q}{\partial a}(\x,a,t) +\delta(a)Q(\x,0,t) \right )d\x\bigg \}da\nonumber \\
\fl &=-\int_0^{\infty}\psi(a) \left [\int_{ \calU}Q(\x,a,t)d\x\right ] da:=-J^{\Psi}(t).
\label{dSocc}
\end{eqnarray}
We have used the divergence theorem and continuity of the the flux across the interface $\partial \calU$.
The term $J^{\Psi}(t)$ is the total absorption flux within the trap and is equivalent to the FPT density. In particular, the MFPT for absorption (assuming it exists) is
\begin{equation}
\tau^{\Psi}:=\E[\calT]=\int_0^{\infty} tJ^{\Psi}(t)dt =-\left . \frac{\partial \J^{\Psi}(s)}{\partial s}\right |_{s=0},
\label{MFPT0}
\end{equation}
where $\J^{\Psi}(s)=\int_0^{\infty}\e^{-st}J^{\Psi}(t)dt$.

One interpretation of a non-exponential occupation time threshold distribution $\Psi(a)$ is that it represents a trap whose reactivity $\kappa$ depends on the amount of time a particle spends within the trap, that is, $\kappa=\kappa(a)$. For example, the reactivity may be a function of some internal state of the particle that is itself dependent on the encounter time between the particle and the trap. In other words,
\begin{equation}
\Psi(a)=\exp\left (-\int_0^a\kappa(a')da'\right ).
\label{intk}
\end{equation}
Multiplying both sides of equations (\ref{Pocca})--(\ref{Poccd}) and integrating with respect to $a$ then gives
\numparts
\begin{eqnarray}
\label{pza}
\fl &\frac{\partial p^{\Psi}(\x,t)}{\partial t}=D\nabla^2 p^{\Psi}(\x,t),\quad \x \in \Omega \backslash \calU,\\
\label{pzb}
\fl &\nabla p^{\Psi}(\x,t) \cdot \n=0, \  \x \in \partial \Omega, \\
\label{pzc}
\fl & \frac{\partial q^{\Psi}(\x,t)}{\partial t}=\overline{D}\nabla^2 q^{\Psi}(\x,t)-\int_0^{\infty}\kappa(a)\Psi(a)\calQ(\x,a,t)da,\quad \x \in \calU  , \\
 \label{pzd}
\fl &p^{\Psi}(\x,t)=q^{\Psi}(\x,t),\quad \nabla Dp^{\Psi}(\x,t)\cdot \n_0 =\overline{D}\nabla q^{\Psi}(\x,t)\cdot \n_0 ,\   \x \in \partial \calU.
\end{eqnarray}
\endnumparts 
An important observation is that this does not yield a closed system of equations for the densities $p^{\Psi}(\x,t)$ and $q^{\Psi}(\x,t)$ due to the final term on the right-hand side of equation (\ref{pzc}). That is, the encounter-based model does not simply consist of replacing a constant reactivity by a time-dependent reactivity $\kappa(t)$, which would result in a term of the form $\kappa(t) q^{\Psi}(\x,t)$.

\section{Occupation time propagator for a reaction-subdiffusion model}

 \subsection{Propagator equation for a CTRW}
 
Consider a CTRW on a $d$-dimensional lattice $\ellb\in \Gamma$ with uniform lattice spacing $\Delta x$. (For the moment, we take the lattice to be infinite.) Let $\ellb(t)$ be the lattice site occupied at time $t$. Waiting times between jump events are independent identically distributed random variables with probability density $u(\tau)$. We also assume that a jump $\ellb'\rightarrow \ellb$ occurs with probability $g_{\ellb\ellb'}$ such that $\sum_{\ellb\in \Gamma}g_{\ellb\ellb'}=1$. Given some sublattice $\Gamma_0\subset \Gamma$, we define the CTRW functional
\begin{equation}
\chi(t)=\int_0^tF(\ellb(\tau))d\tau,\quad F(\ellb)= \sum_{\ellb'\in \Gamma_0}\delta_{\ellb,\ellb'},
\end{equation}
which specifies the amount of time the random walker spends on the sublattice in the time interval $[0,t]$.
Let $P_{\ellb}(a,t)$ denote the joint probability density or propagator for the pair $(\ellb(t),\chi(t))$. It follows that
 \begin{eqnarray}
 \label{A1}
& P_{\ellb}(a,t)=\bigg \langle \delta\left (a-\chi(t) \right )\bigg \rangle_{\ellb(0)=\ellb_0}^{\ellb(t)=\ellb} ,
 \end{eqnarray}
 where expectation is taken with respect to all CTRWs realized by $\ellb(\tau)$ between $\ellb(0)=\ellb_0$ and $\ellb(t)=\ellb$. 
  Introduce the generator
 \begin{eqnarray}
G_{\ellb}(z,t)&=\bigg\langle \exp \left ( -z \chi(t)\right )\bigg \rangle_{\ellb(0)=\ellb_0}^{\ellb(t)=\ellb}=\int_0^{\infty} \e^{-za}P_{\ellb}(a,t)da.
 \end{eqnarray}
Analogous to Brownian functionals, one can derive an evolution for the propagator using a discrete version of a Feynman-Kac formula following along analogous lines to Ref. \cite{Carmi11}.

Let $w_{\ellb}(a,t)dt$ be the probability that the particle jumps to the lattice site $\ellb(t)=\ellb$ with $\chi(t)=a$ in the time interval $[t,t+dt]$. The propagator can be expressed as
\begin{equation}
\label{Pw}
P_{\ellb}(a,t)=\int_0^tU(\tau)w_{\ellb}(a-\tau F(\ellb) ,t-\tau)d\tau, \quad \ell \in \Gamma,
\end{equation}
where $U(\tau)=1-\int_0^{\tau}u(\tau')d\tau'$ is the probability of not jumping in a time interval of length $\tau$.  In order to arrive at $(\ellb,a)$ at time $t$, the particle must have hopped from another site $\ellb'\in \Gamma$ with probability $g_{\ellb\ellb'}$. Assuming that the last jump occurred at time $t-\tau$, we have
\begin{eqnarray}
\fl w_{\ellb}(a,t)&=P_{\ellb}^{(0)}\delta(a)\delta(t)+\sum_{\ellb'\in \Gamma}g_{\ellb\ellb'}\int_0^t u(\tau)w_{\ellb'}(a-\tau F(\ellb'),t-\tau)d\tau,
\end{eqnarray}
with
$P_{\ellb}^{(0)}=P_{\ellb}(a=0,t=0)$.
Laplace transforming with respect to $a$ by setting $H_{\ellb}(z,t)=\int_0^{\infty}\e^{-za}w_{\ellb}(a,t)da$ gives
\begin{eqnarray}
H_{\ellb}(z,t)&=P_{\ellb}^{(0)} \delta(t) +\sum_{\ellb'\in \Gamma} g_{\ellb\ellb'}\int_0^t u(\tau)\e^{-z\tau F(\ellb')}H_{\ellb'}(z,t-\tau)d\tau
\end{eqnarray}
and Laplace transforming the result with respect to $s$ yields
\begin{eqnarray}
\label{wn}
\H_{\ellb}(z,s)&=P^{(0)}_{\ellb} +\sum_{\ell'}\widetilde{u}(s+zF(\ellb')  )g_{\ellb\ellb'}  \H_{\ellb'}(z,s) ,
\end{eqnarray}
where $\widetilde{u}(s)=\int_0^{\infty}\e^{-s\tau} u(\tau)d\tau$. Performing the double Laplace transform of equation (\ref{Pw}) shows that
\begin{eqnarray}
\G_{\ellb}(z,s)&=\widetilde{U}(s+zF(\ellb)) \H_{\ellb}(z,s),
\end{eqnarray}
with $\widetilde{U}(s)=(1-\widetilde{u}(s))/{s}$.
Finally, substituting into equation (\ref{wn}) gives 
\begin{eqnarray}
\fl &(s+zF(\ellb))\widetilde{G}_{\ellb}(z,s)\nonumber \\
\fl &=P_{\ellb}^{(0)} + \left [ \sum_{\ellb'}g_{\ellb\ellb'} \widetilde{v}(s+z F(\ellb'))\widetilde{G}_{\ellb'}(z,s)-\widetilde{v}(s+z F(\ellb))\widetilde{G}_{\ellb}(z,s)\right ],
\label{Pn}
\end{eqnarray}
where $\widetilde{v}(s)={s\widetilde{u}(s)}/(1-\widetilde{u}(s))$.

\subsection{Continuum limit}

Consider an exponential waiting-time density, $u(\tau)=h\e^{-h\tau}$, for which the relevant Laplace transforms are
$\widetilde{u}(s) ={h}/(h+s)$ and $\widetilde{v}(s)=h$.
Equation (\ref{Pn}) then reduces to the form
  \begin{eqnarray}
 (s +zF(\ellb))\G_{\ellb}(z,s)=h\sum_{\ellb'}[g_{\ellb\ellb'}\G_{\ellb'}(z,s)-\G_{\ellb}(z,s) ].
 \label{calG}
\end{eqnarray}
Inverting with respect to $z$ and $s$ leads to a Feynman-Kac equation for a classical random walk with a constant hopping rate $h$:
\begin{eqnarray}
\fl \frac{\partial P_{\ellb}(a,t)}{\partial t}&=h\left [\sum_{\ellb'}g_{\ellb\ellb'} P_{\ellb'}(a,t)- P_{\ellb}(a,t)\right ]
-\left[\frac{\partial P_{\ellb}}{\partial a}(a,t) + \delta(a)P_{\ellb}(0,t)\right ] \sum_{\ellb'\in \Gamma_0} \delta_{\ellb,\ellb'}.\nonumber \\
\fl 
\label{calP}
\end{eqnarray}
We first consider the well-known continuum limit of the simpler equation \cite{Hughes95}
\begin{eqnarray}
 \frac{\partial p_{\ellb}(t)}{\partial t}&=h\left [\sum_{\ellb'}g_{\ellb\ellb'} p_{\ellb'}(t)- p_{\ellb}(a,t)\right ].
\end{eqnarray}
Suppose that $g_{\ellb\ellb'}=g(\ellb-\ellb')$.
Taking discrete Fourier transforms and using the convolution theorem (on infinite lattices) gives
\begin{eqnarray}
\frac{\partial \widehat{p}_{\k}(t)}{\partial t}= h\left [\widehat{g}(\k) \widehat{p}_{\k}(t)- \widehat{p}_{\k}(t)\right ],
\end{eqnarray}
where $\widehat{p}_{\k}(t)=\sum_{\ellb\in \Gamma}\e^{i\k\cdot \ellb}p_{\ellb}(t)$ etc. Setting $\x=\ellb \Delta x$, $\q=\k/\Delta \x$, $p(\x,t)(\Delta x)^d =p_{\ellb}(t)$ and $\widehat{p}(\q,t)=\widehat{p}_{\k/\Delta x}(t)$, we have
\begin{equation}
\frac{\partial \widehat{p}(\q,t)}{\partial t}= h\left [\widehat{g}(\Delta x\q) \widehat{p}(\q,t)- \widehat{p}(\q,t)\right ],
\end{equation}
with
\begin{equation}
\widehat{p}( {\bf q},t)=(\Delta x)^d \sum_{\k\in \Gamma}\e^{i{\bf q} \cdot \x}p(\x,t).
\end{equation}
Assuming that $g(\ellb)$ has finite moments and its Fourier transform has the leading order Taylor series expansion $\widehat{g}(\Delta x\, \q)\approx1-(\Delta x)^2 q^2$ with $q^2=\q\cdot \q$, then
\begin{equation}
\frac{\partial \widehat{p}(\q,t)}{\partial t}\approx - h(\Delta x)^2 q^2 \widehat{p}(\q,t).
\end{equation}
Finally, assuming that the hopping rate is of the form
\begin{equation}
\label{hhbar}
h=\frac{D}{(\Delta x)^2}, \end{equation}
we can take the continuum limit $\Delta x\rightarrow 0$ with $ \widehat{p}(\q,t)$ the Fourier transform of $p(\x,t)$. Noting that $q^2$ is the Fourier transform of the Laplacian $\nabla^2$, we obtain the classical diffusion equation on $\R^d$ with diffusivity $D$. Returning to the propagator equation (\ref{calP}), we can rewrite the second term on the right-hand side in the form
\begin{eqnarray}
\left[\frac{\partial P(\x,a,t)}{\partial a}+ \delta(a)P(\x,0,t)\right ] (\Delta x)^d\sum_{\x' /\Delta x\in \Gamma_0} \frac{\delta_{\x/\Delta x,\x'/\Delta x}}{(\Delta x)^d},
\end{eqnarray}
where $P(\x,a,t)(\Delta x)^d=P_{\ellb}(x,t)$. Taking the continuum limit of equation (\ref{calP}) with $\Gamma_0\rightarrow \calU$ then gives
\begin{eqnarray}
\fl &\frac{\partial P(\x,a,t)}{\partial t}=D\nabla^2 P(\x,a,t)-\left (\frac{\partial P}{\partial a}(\x,a,t) +\delta(a)P(\x,0,t) \right )\int_{\calU}\delta(\x-\y)d\y,\ \x\in \R^d.\nonumber \\
\fl
\end{eqnarray}
This is equivalent to equations (\ref{Pocca}), (\ref{Poccc}) and (\ref{Poccd}) for $\Omega=\R^d$, and $D=\overline{D}$ (after setting $P=Q$ for all $\x\in \calU$).
It is also possible to extend the above formal analysis to the case of a finite lattice $\Gamma$ such that $\Gamma \rightarrow \Omega$ in the continuum limit and a no-flux boundary condition on $\partial \Omega$ by modifying the probability matrix $g_{\ellb\ellb'}$ accordingly. Furthermore, the bulk hopping rates on the lattices $\Gamma\backslash \Gamma_0$ and $\Gamma_0$ could be different so that $D\neq \overline{D}$ in the continuum limit. (However, care has to be taken in defining the hopping rates at the interface between the two lattices).

Following Refs.  \cite{Metzler07,Carmi11}, we now consider the heavy-tailed waiting time density
\begin{equation}
u(\tau)\sim \frac{B_{\alpha}}{|\Gamma(-\alpha)|}\tau^{-(1+\alpha)},\quad 0 < \alpha <1 ,
\end{equation}
where $\Gamma(\mu)$ is the gamma function
 \begin{equation}
 \label{gamf}
\Gamma(\mu)=\int_0^{\infty}\e^{-t}t^{\mu-1}dt,\ \mu\neq 0,-1,-2\ldots.
\end{equation}
The Laplace transform  for small $s$ is 
\begin{equation}
\widetilde{u}(s)\sim 1-B_{\alpha}s^{\alpha} \mbox { and } \widetilde{v}(s)\sim \frac{s}{B_{\alpha}}(s^{-\alpha}-B_{\alpha}),
\end{equation}
where $B_{\alpha}$ plays the role of the inverse of the hopping rate $h$. We also assume that $u(t)\sim h\e^{-ht}$ for small $t$, so that $\widetilde{u}(s)\sim h/s$ and $\widetilde{v}(s)\sim h$ in the limit $s\rightarrow \infty$.
Given the lattice spacing $\Delta x$, we take $h$ to be given by equations (\ref{hhbar}), whereas
\begin{equation}
B_{\alpha}=\frac{(\Delta x)^2}{2D_{\alpha}}
\end{equation}
for a constant $D_{\alpha}$. Setting $\G(\x,z,s)(\Delta x)^d=\G_{\x/\Delta x}(z,s)$
and taking the continuum limit of equation (\ref{Pn}) along similar lines to the classical random walk formally gives
\begin{eqnarray}
\label{Gx0}
\fl (s+zF(\x))\G(\x,z,s)&=P(x,0,0)+D_{\alpha}[s+zF(\x)]^{1-\alpha}\nabla^2 \G(\x,z,s),
\end{eqnarray}
where 
\begin{equation}
 F(\x)=\int_{\calU}\delta(\x-\y)d\y.
\end{equation}
Let $\calG(\x,s)$ be the solution to the simpler equation
\begin{eqnarray}
\label{calGx0}
s\calG(\x,s)&=P(\x,0,0)+D_{\alpha}s^{1-\alpha}\nabla^2 \calG(\x,s).
\end{eqnarray}
Inverting the Laplace transform in $s$ then yields the well-known fractional diffusion equation
\begin{eqnarray}
\label{fracalG}
	&\frac{\partial \calG(x,t)}{\partial t} =D_{\alpha}{\mathcal D}_t^{1-\alpha}\nabla^2 \calG(x,t),
	\end{eqnarray}
where the fractional derivative $ {\mathcal D}_t^{1-\alpha} $ is defined in Laplace space according to \cite{Barkai00}
\begin{equation}
\int_0^{\infty}\e^{-st}  {\mathcal D}_t^{1-\alpha} f(t)dt=s^{1-\alpha} \widetilde{f}(s).
\end{equation}
It can also be written as the fractional Riemann-Liouville equation \cite{Barkai00}
\begin{equation}
 {\mathcal D}_t^{1-\alpha} f(t)=\frac{1}{\Gamma(a)}\frac{\partial}{\partial t}\int_0^t\frac{f(t')}{(t-t')^{1-\alpha}}dt'.
 \end{equation}
 Comparison of equations (\ref{Gx0}) and (\ref{calGx0}) implies that
 \begin{equation}
\G(\x,z,s)=\calG(\x,s+zF(\x)) \Rightarrow  G(\x,z,t)=\e^{-zF(\x)t}\calG(\x,t).
\end{equation}
Finally, substituting for $\calG$ in equation (\ref{calGx0}), we obtain the following Feynman-Kac equation for the generator:
\begin{eqnarray}
\label{fracG}
\fl	&\frac{\partial G(\x,z,t)}{\partial t} =\e^{-zF(\x)t}D_{\alpha}{\mathcal D}_t^{1-\alpha}\e^{zF(\x)t}\nabla^2 G(\x,z,t)-zF(\x) G(\x,z,t),\quad \x\in \R^d.
	\end{eqnarray}
As in the case of the classical random walk, this equation can be generalized to subdiffusion in a bounded domain $\Omega$. Moreover, the waiting time densities on the lattices $\Gamma\backslash \Gamma_0$ and $\Gamma_0$ could also be different, resulting in the distinct fractional diffusion operators $D_{\beta}{\mathcal D}_t^{1-\beta}$ for $\x\in \Omega$ and $D_{\alpha}{\mathcal D}_t^{1-\alpha}$ for $\x\in \calU$.

Mathematically speaking, equation (\ref{fracG}) is identical in form to the reaction-subdiffusion equation obtained from an evanescent CTRW \cite{Yuste07,Yuste10,Abad10}, with $k(\x)=zF(\x)$ corresponding to a position-dependent first-order reactivity. It is convenient to rewrite (\ref{fracG}) in an analogous form to equations (\ref{za}) -- (\ref{zd}) for normal diffusion.  Therefore, setting $G(\x,z,t)=\calP(\x,z,t)$ for $\x \in \Omega\backslash \calU$ and $G(\x,z,t)=\calQ(\x,z,t)$ for $\x \in  \calU$, we find that
\numparts
\begin{eqnarray}
\label{eva}
\fl &\frac{\partial \calP(\x,z,t)}{\partial t}=D_{\beta}{\mathcal D}_t^{1-\beta}\nabla^2 \calP(\x,z,t),\quad \x \in \Omega \backslash \calU,\\
\label{evb}
\fl &D_{\beta}{\mathcal D}_t^{1-\beta} \nabla \calP(\x,z,t) \cdot \n=0, \  \x \in \partial \Omega, \\
\label{evc}
\fl & \frac{\partial \calQ(\x,z,t)}{\partial t}=\e^{-z t}D_{\alpha}{\mathcal D}_t^{1-\alpha}\e^{z t}\nabla^2 \calQ(\x,z,t)-z\calQ(\x,z,t),\quad \x \in \calU  , \\
 \label{evd}
\fl &\calP(\x,z,t)=\calQ(\x,z,t),\   \x \in \partial \calU\\
\fl& D_{\beta}{\mathcal D}_t^{1-\beta}\nabla \calP(\x,z,t)\cdot \n_0 =\e^{-zt}D_{\alpha}{\mathcal D}_t^{1-\alpha}\e^{zt}\nabla \calQ(\x,z,t) \cdot \n_0 ,\   \x \in \partial \calU.
\label{eve}
\end{eqnarray}
\endnumparts 
Within the context of our encounter-based approach, the solutions $\calP(\x,z,t)$ and $\calQ(\x,z,t)$ are the Laplace transforms of the occupation time propagators $P(\x,a,t)$ and $Q(\x,a,t)$ in the domains $\x\in \Omega\backslash \calU$ and $\calU$, respectively. The propagators can then be used to determine the corresponding marginal probability densities $p^{\Psi}$ and $q^{\Psi}$ according to equations (\ref{peep}) and (\ref{qeep}).

\section{FPT problem for a spherical trap}

One major potential application of our encounter-based reaction-subdiffusion model is to neurotransmitter receptor trafficking in neurons. Driven by advances in single particle tracking (SPT), it has been established that the diffusion-trapping of protein receptors in post-synaptic regions of the cell membrane plays a major role in determining the strength of synaptic connections between neurons (see the recent review \cite{Triller22} and references therein). These connections are thought to be the molecular substrate of learning and memory. One characteristic feature of post-synaptic regions is that they are packed with scaffolding proteins and other molecular structures that impede the diffusion of receptors. This results in anomalous subdiffusion over a range of timescales. In light of this example, we consider the particular problem of a particle undergoing normal diffusion in $\Omega\backslash \calU$ and subdiffusion in $\calU$. As a further simplification, we will take $\calU$ and $\Omega$ to be concentric $d$-dimensional spheres so that we can exploit spherical symmetry. (For inhibitory synapses located on the cell body of neuron, one could approximate $\calU$ by a disk, for example.) We are interested in determining under what conditions the MFPT for absorption within $\calU$ is finite. (In the case of receptor trafficking, absorption could correspond to internalization of a receptor with the interior of a cell, a process known as endocytosis.) This is very distinct from the scenario shown in Fig \ref{fig1}(a) where a particle subdiffuses within $\Omega\backslash \calU$ until it is absorbed at the boundary $\calU$. Now subdiffusion results in an infinite MFPT and the long-time behavior of the FPT density is characterized by power-law decay \cite{PCBI}.

\subsection{Derivation of the FPT density} Consider the survival probability defined in equation (\ref{Socc}).
Differentiating both sides with respect to $t$ and using equations (\ref{eva})--(\ref{eve}) with $\beta=0$ and $D_0=D$ gives
\begin{eqnarray}
\fl \frac{dS^{\Psi}(t)}{dt}&= \int_{\Omega\backslash \calU} \frac{\partial p^{\Psi}(\x,t)}{\partial t}d\x+\int_{ \calU}  \frac{\partial q^{\Psi}(\x,t)}{\partial t}d\x\nonumber \\
\fl &= \int_0^{\infty}\Psi(a) {\mathcal L}_a^{-1} \bigg\{ \int_{\Omega\backslash \calU} \frac{\partial \calP (\x,z,t)}{\partial t}d\x+\int_{ \calU}  \frac{\partial \calQ(\x,z,t)}{\partial t}d\x\bigg \}da\nonumber \\
\fl &=\int_0^{\infty}\Psi(a){\mathcal L}_a^{-1}\bigg \{\int_{\Omega\backslash \calU}  D\nabla^2 \calP(\x,z,t)d\x+
\e^{-zt}D_{\alpha} {\mathcal D}_t^{1-\alpha}\e^{zt}\int_{ \calU} \nabla^2 \calQ(\x,z,t)d\x\nonumber \\
\fl &\hspace{3cm} -z\int_{ \calU}\calQ(\x,z,t)d\x\bigg \}da\nonumber \\
\fl &=-\int_0^{\infty}\psi(a) \left [\int_{ \calU}Q(\x,a,t)d\x\right ] da:=-J^{\Psi}(t).
\label{dSocc2}
\end{eqnarray}
Thus the probability flux $J^{\Psi}(t)$ is related to the propagator $Q(\x,a,t)$ in an identical fashion to that of normal diffusion. Moreover, the flux determines the MFPT $\tau^{\Psi}$ according to equation (\ref{MFPT0}). The latter involves the Laplace transformed flux $\J^{\Psi}(s)$ which is easier to calculate. In particular,
\begin{equation}
\label{jag}
\J^{\Psi}(s)=\int_0^{\infty} \psi(a)  {\mathcal L}^{-1}_{a}\left [\int_{ \calU}  \QQ(\x,z,s)d\x\right ]da,
\end{equation}
with 
\numparts
\begin{eqnarray}
\label{evaLT}
\fl &D\nabla^2 \PP(\x,z,s)-s \PP(\x,z,s)=-\delta(\x-\x_0),\quad \x \in \Omega \backslash \calU,\\
\label{evbLT}
\fl &\nabla \PP(\x,z,s) \cdot \n=0, \  \x \in \partial \Omega, \\
\label{evcLT}
\fl &D_{\alpha}[s+z]^{1-\alpha}\nabla^2 \QQ(\x,z,s)- (s+z)\QQ(\x,z,s)=0,\quad \x \in \calU  , \\
 \label{evdLT}
\fl &\PP(\x,z,s)=\QQ(\x,z,s),\   \x \in \partial \calU\\
\fl& D\nabla \PP(\x,z,s)\cdot \n_0 = D_{\alpha} (s+z)^{1-\alpha}\nabla \QQ(\x,z,s) \cdot \n_0 ,\   \x \in \partial \calU.
\label{eveLT}
\end{eqnarray}
\endnumparts 

One geometrical configuration where the BVP given by equations (\ref{evaLT})--(\ref{eveLT}) can be solved explicitly is for a pair of concentric $d$-dimensional spheres: $\Omega =\{\x\in \R^d\,|\, 0\leq  |\x| <R_2\}$ and $\calU =\{\x\in \R^d\,|\, 0\leq  |\x| <R_1\}$, with $0<R_1<R_2$.
We also assume that the initial distribution of the particle is spherically symmetric, that is, $\PP(\x,z,0)=\delta(|\x|-r_0)/\Omega_d r_0^{d-1}$, where $\Omega_d$ is the surface area of the unit sphere in $\R^d$ and $R_1<r_0<R_2$. This allows us to exploit spherical symmetry by setting $\PP(\x,z,t)=\PP(r,z,t)$ and $\QQ(\x,z,t)=\QQ(r,z,t)$ with $r=|\x|$. Rewriting equations (\ref{evaLT})--(\ref{eveLT}) in terms of spherical polar coordinates gives
  \numparts
\begin{eqnarray}
\label{LTospha}
 \fl   &D\frac{\partial^2\PP}{\partial r^2} + D\frac{d - 1}{r}\frac{\partial \PP}{\partial r} -s\calP(r,z, s) = -\frac{1}{\Omega_d r_0^{d - 1}} \delta(r - r_0), \ R_1<r <R_2,\\ 
 \fl & \left . \frac{\partial }{\partial r}\PP(r,z,s)\right |_{r=R_2}=0,
  \label{LTosphb}\\
\fl   &D_{\alpha}\frac{\partial^2\QQ}{\partial r^2} + D_{\alpha}\frac{d - 1}{r}\frac{\partial \QQ}{\partial r} -(s+z)^{\alpha}\QQ(r,z, s) =0,\quad 0<r < R_1,
 \label{LTosphc}\\
\fl & \PP(r,z,s)=\QQ(r,z,s),\ D\left . \frac{\partial }{\partial r}\PP(r,z,s)\right |_{r=R_1}=D_{\alpha}(s+z)^{1-\alpha}\left . \frac{\partial }{\partial r}\QQ(r,z,s)\right |_{r=R_1}.
 \label{LTosphd}
\end{eqnarray}
\endnumparts
Note that in the special case $\alpha=0$ (normal diffusion everywhere) we recover the BVP analyzed in Ref. \cite{Bressloff22}. The latter was solved in terms of modified Bessel functions and we can carry over the analysis to the subdiffusive case with minor modifications.

The general solution of equation (\ref{LTospha}) for $R_1<r <R_2$ is
  \begin{eqnarray}
\label{cP0}
 \fl    \PP(r,z, s) = {B}(z,s) r^\nu I_\nu(\sqrt{s/D} r)  + {C}(z,s)r^\nu K_\nu(\sqrt{s/D} r) + G_{\rm mh}(r, s| r_0)  ,
\end{eqnarray}
with $\nu = 1 - d/2$. In addition, $I_{\nu}$ and $K_{\nu}$ are modified Bessel functions of the first and second kind, respectively. 
The first two terms on the right-hand side of equation (\ref{cP0}) are the solutions to the homogeneous version of equation (\ref{LTospha}) and $G_{\rm mh}$ is the modified Helmholtz Green's function satisfying
  \numparts
\begin{eqnarray}
\label{Ga}
 \fl   &D\frac{\partial^2G_{\rm mh}}{\partial r^2} + D\frac{d - 1}{r}\frac{\partial G_{\rm mh}}{\partial r} -sG_{\rm mh}  = -\frac{1}{\Omega_d r_0^{d - 1}} \delta(r- r_0), \ R_1<r <R_2,\\ 
 \fl & G_{\rm mh}(R_1,s|r_0)=0,\quad \left .\frac{\partial }{\partial r}G_{\rm mh}(r,s|r_0)\right |_{r=R_2}=0.
 \label{Gb}
\end{eqnarray}
\endnumparts
One finds that \cite{Redner01}
\begin{eqnarray}
\label{GG}
  G_{\rm mh}(r, s| r_0) = \frac{ (rr_0)^\nu }{D\Omega_d}\frac{C_{\nu}(r_{<},R_1;s)\overline{C}_{\nu}(r_{>},R_2;s)}{\overline{C}_{\nu}(R_1,R_2;s)},
\end{eqnarray}
where $r_< = \min{(r, r_0)}$, $r_> = \max{(r, r_0)}$, and
\numparts
\begin{eqnarray}
\fl  C_{\nu}(a,b;s)&=I_{\nu}(\sqrt{s/D} a)K_{\nu}(\sqrt{s/D} b)-I_{\nu}(\sqrt{s/D} b)K_{\nu}(\sqrt{s/D} a),\\
 \fl \overline{C}_{\nu}(a,b;s)&=I_{\nu}(\sqrt{s/D} a)K_{\nu-1}(\sqrt{s/D} b)+I_{\nu-1}(\sqrt{s/D} b)K_{\nu}(\sqrt{s/D}a).
\end{eqnarray}
\endnumparts
Also note that
\begin{equation}
\widetilde{J}_{\infty}(s):=4\pi R_1^2\left .\frac{\partial }{\partial r}G_{\rm mh}(r,s|r_0)\right |_{r=R_1}=\left (\frac{r_0}{R_1}\right )^{\nu} \frac{\overline{C}_{\nu}(r_0,R_2;s)}{\overline{C}_{\nu}(R_1,R_2;s)}
\end{equation}
is the corresponding flux into a totally absorbing surface $\partial \calU $.
Similarly, the homogeneous equation (\ref{LTosphc}) has the solution
 \begin{eqnarray}
 \label{cQ0}
  \fl  \QQ(r,z, s) = \widehat{B}(z,s) r^\nu I_\nu(\sqrt{(s+z)^{\alpha}/D_{\alpha}} r)  + \widehat{C}(z,s)r^\nu K_\nu(\sqrt{(s+z)^{\alpha}/D_{\alpha}}r)
\end{eqnarray}
for $0<r < R_1$.
There are four unknown coefficients but only one boundary condition (\ref{LTosphb}) and two continuity conditions, see equations (\ref{LTosphd}). The fourth condition is obtained by requiring that the solution remains finite at $r=0$. The details of the latter will depend on the dimension $d$. 

\subsection{The 3D sphere ($d=3$)} In the 3D case, equations (\ref{cP0}) and (\ref{cQ0}) become
 \begin{eqnarray}
 \label{cP}
 \fl    \PP(r,z, s) = {B}(z,s) \sqrt{\frac{2}{\pi \theta}}\frac{\cosh \theta r} {r} + {C}(z,s)\sqrt{\frac{\pi}{2 \theta}}\frac{\e^{-\theta r}}{r}+ G_{\rm mh}(r, s| r_0) 
\end{eqnarray}
for $R_1<r<R_2$ and
 \begin{eqnarray}
 \label{cQ}
    \QQ(r,z, s) =E(z,s)\frac{\sinh \theta_{\alpha} r}{r}, \ 0<r < R_1.
\end{eqnarray}
We have introduced the variables
\begin{equation}
\theta=\sqrt{\frac{s}{D}},\quad \theta_{\alpha}=\sqrt{\frac{(s+z)^{\alpha}}{D_{\alpha}}}.
\end{equation}
The 3D case with $\alpha=1$ and $D_{\alpha}=D$ was analyzed in Ref. \cite{Bressloff22}. For $\alpha >0$, the reflecting boundary condition at $r=R_2$ implies that
\begin{equation}
B(z,s)= \frac{\pi}{\Lambda(\theta R_2)}C(z,s),\quad \Lambda(y)=\frac{[y-1]\e^{2y}}{y+1} -1.
\end{equation}
Substituting equation (\ref{cQ}) into equation (\ref{jag}), after rewriting the latter in spherical polar coordinates, shows that 
\begin{eqnarray}
\J^{\Psi}(s)&=4\pi \int_0^{\infty}\psi(a) {\mathcal L}_a^{-1}\left [E(z,s)\int_{0}^{R_1}r \sinh (\theta_{\alpha} r ) dr\right ]da.
\label{bb}
\end{eqnarray}
Evaluating the integral with respect to $r$,
\begin{equation}
\int_0^{R_1}r \sinh(\theta_{\alpha} r) dr= \frac{1}{{\theta_{\alpha}}^2} \left (\theta_{\alpha} R_1 \cosh (\theta_{\alpha} R_1)-\sinh (\theta_{\alpha} R_1)\right ),
\end{equation}
 and using the solution for $E(z,s)$ obtained by imposing the continuity conditions at $r=R_1$, we find that
 \begin{eqnarray}
\J^{\Psi}(s)&=\widetilde{J}_{\infty}(s)\int_0^{\infty}\psi(a) {\mathcal L}_a^{-1}\left [\frac{1}{\theta_{\alpha}^2}\widehat{\Lambda}_{\theta,\theta_{\alpha}}(R_1,R_2)  \right ]da,
\label{Jbb}
\end{eqnarray}
where
\begin{eqnarray}
\fl &\widehat{\Lambda}_{\theta,\theta_{\alpha}}(R_1,R_2)=\frac{1}{D_{\alpha}(s+z)^{1-\alpha}- D \calF_1 (\theta_{\alpha} R_1)\calF_2 (\theta R_1,\theta R_2) }. 
\label{Jbb2}
\end{eqnarray}
with
\begin{eqnarray}
\fl \calF_1(x)=\frac{\sinh x}{x \cosh x-\sinh x},\quad \calF_2(x,y)=\frac{2\left (x\sinh (x) - \cosh (x) \right )-\Lambda(y)\e^{-x} \left (1+x\right )}{2\cosh( x )+ \Lambda(y)\e^{-x}}.\nonumber \\
\fl
\end{eqnarray}
Substituting equation (\ref{bb}) into (\ref{MFPT0}) gives
\begin{eqnarray}
\fl  \tau^{\Psi}&=-\left .\frac{\partial}{\partial s}\widetilde{J}^{\Psi}(s)\right |_{s=0}=\tau_{\infty}- \left .\frac{\partial}{\partial s}\right |_{s=0}\int_0^{\infty}\psi(a) {\mathcal L}_a^{-1}\left [\frac{1}{\theta_{\alpha}^2}\widehat{\Lambda}_{\theta,\theta_{\alpha}}(R_1,R_2) \right ]da\nonumber\\
\fl &=\tau_{\infty}- \lim_{s\rightarrow 0} \int_0^{\infty}\psi(a) {\mathcal L}_a^{-1}\left [\frac{1}{2\sqrt{sD}}\frac{d}{d\theta}+\frac{d}{dz}\right ] \left [\frac{1}{\theta_{\alpha}^2}\widehat{\Lambda}_{\theta,\theta_{\alpha}}(R_1,R_2) \right ]da,
\end{eqnarray}
where $\tau_{\infty}$ is the MFPT in the case of a totally absorbing surface $\partial \calU$.

In order to calculate the MFPT for absorption, we need to determine the small-$s$ behavior of $\J^{\Psi}(s)$. First note that if $\theta >0$, then as $R_2\rightarrow \infty$ (unbounded domain $\Omega = \R^3$), and
\begin{equation}
\lim_{R_2\rightarrow \infty} \calF_2 (\theta R_1,\theta R_2)=\frac{1}{R_1}+\theta.
\end{equation}
On the other hand, for finite $R_2$, Taylor expanding $ \calF_2 (\theta R_1,\theta R_2)$ with respect to $\theta=\sqrt{s/D}$, we find \cite{Bressloff22}
\begin{equation}
\label{aexp}
 \calF_2 (\theta R_1,\theta R_2) \sim \theta^2\frac{R_2^3-R_1^3}{3R_1}
\end{equation}
and, hence,
\begin{equation}
\fl \widehat{\Lambda}_{\theta,\theta_{\alpha}}(R_1,R_2)\sim \frac{1}{D_{\alpha}(s+z)^{1-\alpha}}+\frac{D}{[D_{\alpha}(s+z)^{1-\alpha}]^2}\calF_1 (\theta_{\alpha} R_1)\theta^2\frac{R_2^3-R_1^3}{3R_1}.
\end{equation}
It follows that
\begin{eqnarray}
\fl \frac{d}{d\theta}\widehat{\Lambda}_{\theta,\theta_{\alpha}}(R_1,R_2)&
 &\sim -\frac{D}{[D_{\alpha}(s+z)^{1-\alpha}]^2}\calF_1 (\theta_{\alpha} R_1)\frac{2\theta(R_2^3-R_1^3)}{3R_1}+O(\theta^2).
\end{eqnarray}
and
\begin{eqnarray}
\fl \frac{d}{dz}\left (\frac{1}{\theta_{\alpha}^2}\widehat{\Lambda}_{\theta,\theta_{\alpha}}(R_1,R_2)\right )& \sim -\frac{1}{(s+z)^2}+O(\theta^2).
\end{eqnarray}
Hence,
\begin{eqnarray}
\fl  \tau^{\Psi}&=\tau_{\infty}+ \int_0^{\infty}\psi(a) {\mathcal L}_a^{-1}\bigg [\frac{1}{z^2}+\frac{1}{zD_{\alpha}z^{1-\alpha}}\calF_1 (\sqrt{z^{\alpha}/D_{\alpha}} R_1)\frac{R_2^3-R_1^3}{3R_1}  \bigg ] da \nonumber \\
\fl &=\tau_{\infty}+\E[a]+\frac{1}{D_{\alpha}}\frac{R_2^3-R_1^3}{3R_1} \int_0^{\infty}\psi(a) {\mathcal L}_a^{-1}\left [ \frac{1}{z^{2-\alpha}}\calF_1 (\sqrt{z^{\alpha}/D_{\alpha}} R_1)\right ]da.
\label{tau3D}
\end{eqnarray}
Note that $\tau^{\Psi}=\E[a]$ when $R_2=R_1$ since the particle spends all of its time within the trap. 

\subsection{The 2D disk ($d=2$)} In the case of concentric disks, equations (\ref{cP0}) and (\ref{cQ0}) become
 \begin{eqnarray}
 \label{2dcP}
   \PP(r,z, s) = {B}(z,s) I_0(\theta r)+ {C}(z,s)K_0(\theta r)+ G_{\rm mh}(r, s| r_0) 
\end{eqnarray}
for $R_1<r<R_2$ and
 \begin{eqnarray}
 \label{2dcQ}
    \QQ(r,z, s) =E(z,s)I_0(\theta_{\alpha} r), \ 0<r < R_1.
\end{eqnarray}
The boundary condition at $r=R_2$ gives
\begin{equation}
B(z,s)=-\frac{K'_0(\theta R_2)}{I_0'(\theta R_2)}C(z,s)=\frac{K_1(\theta R_2)}{I_1(\theta R_2)}C(z,s),
\end{equation}
after using standard Bessel function identities. Substituting equation (\ref{2dcQ}) into equation (\ref{jag}), and using polar coordinates shows that 
\begin{eqnarray}
\J^{\Psi}(s)&=2\pi \int_0^{\infty}\psi(a) {\mathcal L}_a^{-1}\left [E(z,s)\int_{0}^{R_1}r I_0(\theta r)dr\right ]da.
\label{2Dbb}
\end{eqnarray}
Evaluating the integral with respect to $r$,
\begin{equation}
\int_0^{R_1}r I_0(\theta_{\alpha} r) dr= \frac{\theta_{\alpha}R_1}{\theta_{\alpha}^2}I_1(\theta_{\alpha} R_1),
\end{equation}
 and solving for $E(z,s)$ by imposing the continuity conditions at $r=R_1$, we obtain equations (\ref{Jbb})--(\ref{Jbb2}) with $\calF_{1,2}\rightarrow \calH_{1,2}$ with
 \begin{eqnarray}
\calH_1(x)=\frac{I_0(x)}{x I_1(x)},\quad \calH_2(x,y)=x\frac
{I_1(x)K_1(y)-K_1(x) I_1(y)}{I_0(x)K_1(y)+K_0(x)I_1(y)}.
\fl
\end{eqnarray}
Taylor expanding $ \calH_2 (\theta R_1,\theta R_2)$ with respect to $\theta=\sqrt{s/D}$ with
\begin{equation}
I_0(x)\sim 1+\frac{x^2}{2}, \quad I_1(x)\sim \frac{x}{2},\quad k_0(x)\sim -\ln x,\quad K_1(x)\sim \frac{1}{x},
\end{equation}
yields
\begin{equation}
\calH_2(x,y)\sim \theta^2 \frac{R_2^2-R_1^2}{2}.
\end{equation}
Using similar arguments to the 3D case, we obtain the result
\begin{eqnarray}
\fl  \tau^{\Psi} &=\tau_{\infty}+\E[a]+\frac{1}{D_{\alpha}}\frac{R_2^2-R_1^2}{2} \int_0^{\infty}\psi(a) {\mathcal L}_a^{-1}\left [ \frac{1}{z^{2-\alpha}}\calH_1 (\sqrt{z^{\alpha}/D_{\alpha}} R_1)\right ]da.
\label{tau2D}
\end{eqnarray}

\subsection{Conditions on the density $\psi$ for a finite MFPT}

Equations (\ref{tau3D}) and (\ref{tau2D}) can now be used to identify conditions on the occupation time threshold density $\psi$ that result in a finite MFPT for absorption. The first condition is that $\psi$ has a finite first moment, that is, $\E[a]<\infty$. This is independent of the geometry of the domains $\Omega$ and $\calU$ and the properties of subdiffusion. The same condition has been obtained previously for normal diffusion \cite{Bressloff22}. The second condition is that the final integral terms in equations (\ref{tau3D}) and (\ref{tau2D}) are also finite. These terms clearly depend on the geometry and the details of the subdiffusive process as specified by the index $\alpha$, $0<\alpha <1$. Given the requirement $\E[a]> \infty$, we assume that $\psi(a)$ decays exponentially for large $a$. Furthermore, suppose that the dominant contribution to the integral with respect to $a$ occurs in the regime $0<a<a_c$ with $a_c^{\alpha}\ll R_1^2/D_{\alpha}$. We can then use a large-$z$ approximation for the Laplace transform. In particular,
\numparts
\begin{eqnarray}
\label{F1}
&\frac{1}{z^{2-\alpha}}\calF_1 (\sqrt{z^{\alpha}/D_{\alpha}} R_1)\sim \frac{\sqrt{D_{\alpha}}}{z^{2-\alpha/2}R_1},\\
& \frac{1}{z^{2-\alpha}}\calH_1 (\sqrt{z^{\alpha}/D_{\alpha}} R_1)\sim \frac{1}{z^{2-\alpha}}\left [1+\frac{2D_{\alpha}}{R_1^2z^{\alpha}}\right ].
\label{H1}
\end{eqnarray}
\endnumparts
\numparts
Substituting the approximation (\ref{F1}) into equation (\ref{tau3D}) gives
\begin{eqnarray}
 \tau^{\Psi} &\sim \tau_{\infty}+\E[a]+\frac{1}{\sqrt{D_{\alpha}}}\frac{R_2^3-R_1^3}{3R_1^2} \int_0^{\infty}\psi(a) {\mathcal L}_a^{-1}\left [ \frac{1}{z^{2-\alpha/2}}\right ]da\nonumber \\
 &=\tau_{\infty}+\E[a]+\frac{1}{\Gamma(2-\alpha/2)\sqrt{D_{\alpha}}}\frac{R_2^3-R_1^3}{3R_1^2} \E[a^{1-\alpha/2}] \quad \mbox{(3D)}.
 \label{tau3D2}
\end{eqnarray}
Similarly, substituting the approximation (\ref{H1}) into equation (\ref{tau2D}) yields
\begin{eqnarray}
 \tau^{\Psi} &\sim\tau_{\infty}+\E[a]+\frac{1}{D_{\alpha}}\frac{R_2^2-R_1^2}{2} \int_0^{\infty}\psi(a) {\mathcal L}_a^{-1}\left [ \frac{1}{z^{2-\alpha}}\right ]da\nonumber \\
&=\tau_{\infty}+\E[a]+\frac{1}{\Gamma(2-\alpha)D_{\alpha}}\frac{R_2^2-R_1^2}{2}\E[a^{1-\alpha}] \quad \mbox{(2D)}.
\label{tau2D2}\end{eqnarray}
\endnumparts
Hence, for sufficiently large trap radius $R_1$ we have the additional constraints $\E[a^{1-\alpha/2}]<\infty$ in 3D and $\E[a^{1-\alpha}]<\infty$ in $2D$.

\begin{figure}[t!]
\raggedleft
\includegraphics[width=10cm]{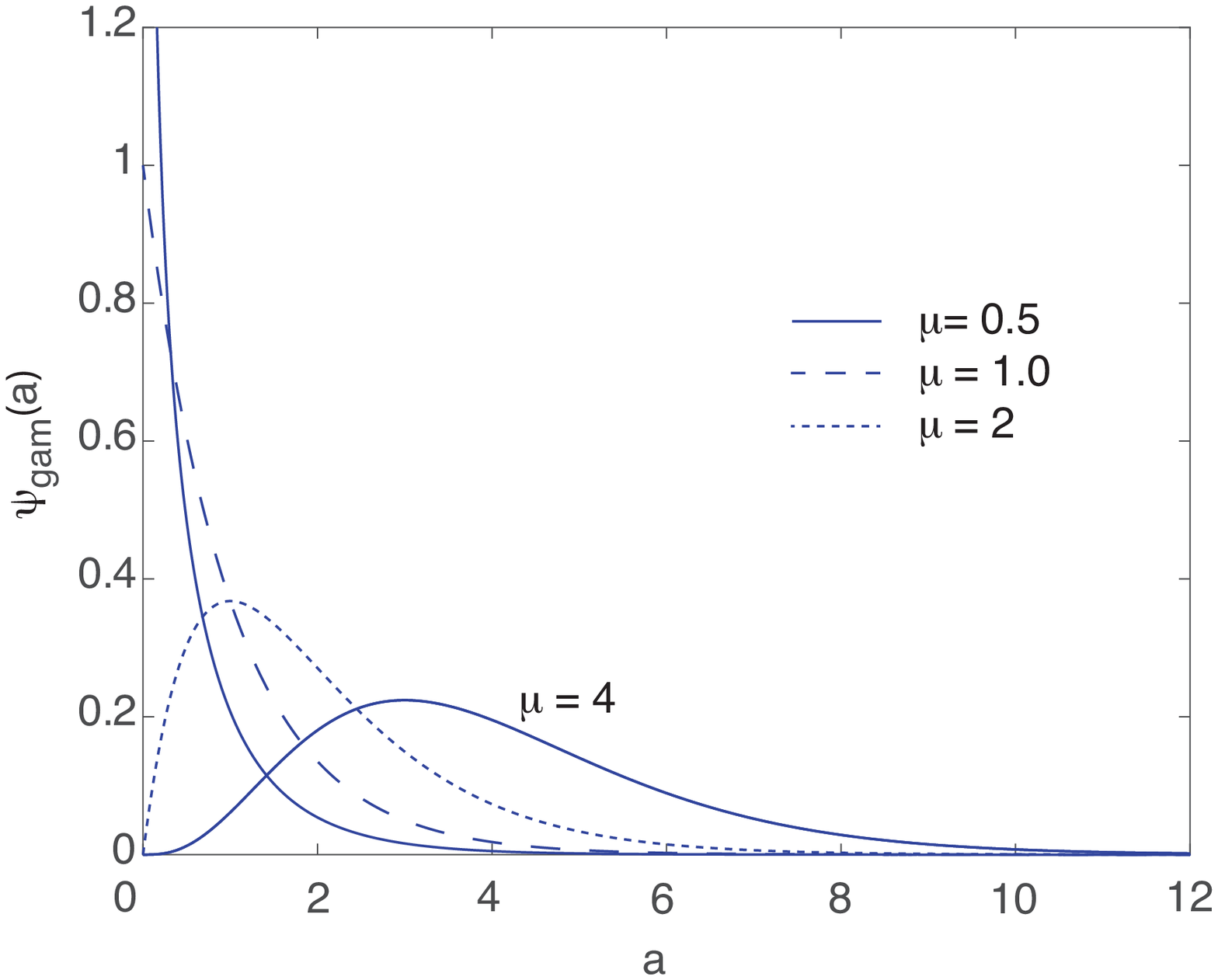} 
\caption{Three-dimensional spherical trap. Plot of MFPT factor $\calM_{\rm gam}(\alpha)$ given by equation (\ref{calM3D}) as a function of the subdiffusion index $\alpha$ for different values of the gamma distribution parameter $\mu$. Subdiffusion reduces (increases) the MFPT compared to normal diffusion  when $\mu<1$ ($\mu>1$.}
\label{fig2}
\end{figure}

 \begin{figure}[b!]
\raggedleft
\includegraphics[width=10cm]{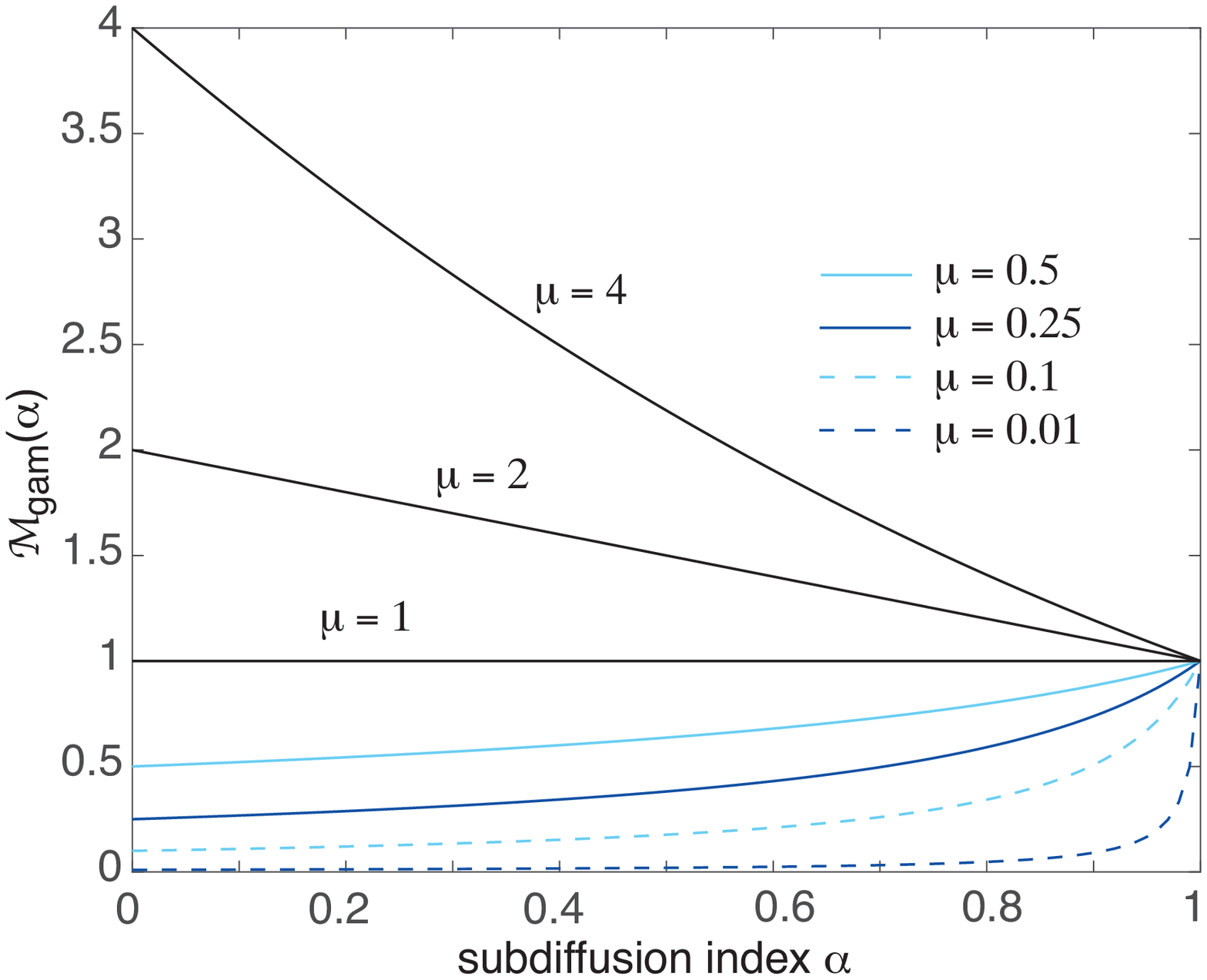} 
\caption{Three-dimensional spherical trap. Plot of MFPT factor $\calM_{\rm gam}(\alpha)$ given by equation (\ref{calM3D}) as a function of the subdiffusion index $\alpha$ for different values of the gamma distribution parameter $\mu$. Subdiffusion reduces (increases) the MFPT compared to normal diffusion  when $\mu<1$ ($\mu>1$.}
\label{fig3}
\end{figure}

One example of a threshold density $\psi(a)$ with finite moments at all integer orders is the gamma distribution, see Fig. \ref{fig2},
\begin{equation}
\label{psigam}
\psi_{\rm gam}(a)=\frac{\kappa_0(\kappa_0 a)^{\mu-1}\e^{-\kappa_0 a}}{\Gamma(\mu)} ,
\end{equation}
where $\kappa_0,\mu$ are positive constants. Note that $\psi(a)$ decays exponentially for large $a$. Indeed, the special case $\mu=1$ corresponds to the exponential distribution with constant reactivity $\kappa_0$. For $\mu \neq 1$ we can use equation (\ref{intk}) to rewrite $\psi(a)$ as
\begin{equation}
\fl \psi_{\rm gam}(a)=\kappa(a)\exp \left (-\int_0^{a}\kappa(a')da'\right ),\quad \kappa(a)=\kappa_0 \frac{ (\kappa_0a)^{\mu-1}\e^{-\kappa_0 a}}{\Gamma(\mu,\kappa_0 a)},
\end{equation}
where $\Gamma(\mu,z)$ is the upper incomplete gamma function:
\begin{equation}
 \Gamma(\mu,z)=\int_z^{\infty}\e^{-t}t^{\mu-1}dt,\ \mu >0.
\end{equation}
For $\mu>1$ one finds that $\kappa(a)\approx 0$ for small $a$ but $\kappa(a)$ increases monotonically with $a$ until it reaches a constant level for large $a$. This could represent a trap that is initially inactive, but becomes more activated as the particle contact time increases. On the other hand, when $\mu < 1$, $\kappa(a)$ is initially large, but reduces to a lower constant level after a sufficient particle contact time. For any positive constant $l$ we have
\begin{eqnarray}
\E[a^{l}]&=\int_0^{\infty} a^{l}\frac{(\kappa_0 a)^{\mu-1}\e^{-\kappa_0 a}}{\Gamma(\mu)} \kappa_0 da\nonumber \\
&=\frac{\kappa_0^{-l}}{\Gamma(\mu)}\int_0^{\infty} y^{l+\mu -1}\e^{-y}dy =\frac{\kappa_0^{-l}\Gamma(\mu+l)}{\Gamma(\mu)}.
\label{mom}
\end{eqnarray}
It follows that
\begin{eqnarray}
\E[a]=\frac{\mu}{\kappa_0},\quad \E[a^2]= \frac{\mu(\mu+1)}{\kappa_0^2},
\end{eqnarray}
On the other hand, setting $l=1-\alpha/2$ in equation (\ref{mom}) and substituting into 
equation (\ref{tau3D2}) gives
\numparts
\begin{eqnarray}
\fl  \tau_{\rm gam} &\sim \tau_{\infty}+\frac{\mu}{\kappa_0}+\frac{\sqrt{\kappa_0}}{\sqrt{D_{\alpha}\kappa_0^{1-\alpha}}}\frac{\Gamma(\mu+1-\alpha/2)}{\Gamma(\mu)\Gamma(2-\alpha/2)}\frac{R_2^3-R_1^3}{3R_1^2} \quad \mbox{(3D)}.
 \label{tau3D2}
\end{eqnarray}
Similarly, setting $l=1-\alpha$ in equation (\ref{mom}) and substituting into 
equation (\ref{tau2D2}) yields 
\begin{eqnarray}
 \tau_{\rm gam}
&=\tau_{\infty}+\frac{\mu}{\kappa_0}+\frac{1}{D_{\alpha} \kappa_0^{1-\alpha}}\frac{\Gamma(\mu+1-\alpha)}{\Gamma(\mu)\Gamma(2-\alpha)}\frac{R_2^2-R_1^2}{2} \quad \mbox{(2D)}.
\end{eqnarray}
\endnumparts

 \begin{figure}[t!]
\raggedleft
\includegraphics[width=10cm]{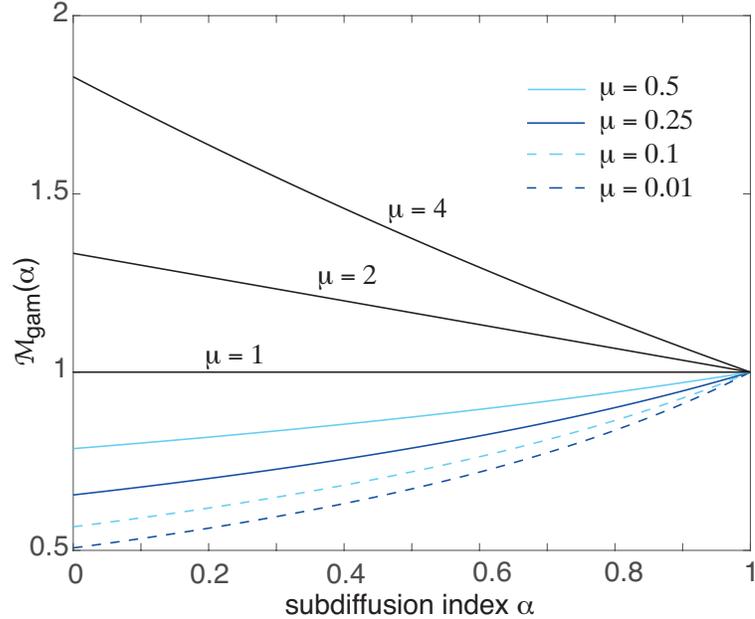} 
\caption{Same as Fig. \ref{fig3} in the case of a 2D circular trap with $\calM_{\rm gam}(\alpha)$ given by equation (\ref{calM2D}).}
\label{fig4}
\end{figure}

We conclude that under the given approximation, the MFPT for absorption is finite. It is then interesting to compare the MFPT for subdiffusion ($\alpha <1$) with the corresponding MFPT for normal diffusion ($\alpha =1)$. Therefore, we introduce the quantity
\begin{equation}
\calM_{\rm gam}(\alpha):=\frac{\Delta \tau_{\rm gam}(\alpha)}{\Delta \tau_{\rm gam}(1)},\quad \Delta \tau_{\rm gam}=\tau_{\rm gam}-\tau_{\infty}-\frac{\mu}{\kappa_0}.
\end{equation}
It follows that
\numparts
\begin{eqnarray}
\calM_{\rm gam}(\alpha)=\frac{\sqrt{D}}{\sqrt{D_{\alpha}\kappa_0^{1-\alpha}}}\frac{\Gamma(\mu+1-\alpha/2)\Gamma(2-1/2)}{\Gamma(\mu+1/2)\Gamma(2-\alpha/2)} \quad \mbox{(3D)}
\label{calM3D}
\end{eqnarray}
and
\begin{eqnarray}
\calM_{\rm gam}(\alpha)=\frac{D}{ D_{\alpha}\kappa_0^{1-\alpha}}\frac{\Gamma(\mu+1-\alpha)}{\Gamma(\mu )\Gamma(2-\alpha)} \quad \mbox{(2D)}.
\label{calM2D}
\end{eqnarray}
\endnumparts
There are a number of general features that emerge from our analysis. 
\medskip

\noindent (i) The effective diffusion coefficient for $\alpha <1$ is $D_{\alpha}\kappa_0^{\alpha -1}$. This combines the subdiffusion parameter $D_{\alpha}$ with the parameter $\kappa_0$ that determines the mean reactivity. The entanglement of the two processes is consistent with previous studies of reaction-subdiffusion models with evanescence \cite{Yuste07,Yuste10,Abad10}. 
\medskip

\noindent (ii) Suppose $D=D_{\alpha}\kappa_0^{1-\alpha}$ so that the effective diffusion coefficients of subdiffusion and diffusion are identical. The scale factor $\calM_{\rm gam}(\alpha)\neq 1$ for $\alpha <1$, which means that subdiffusion still affects the MFPT.
\medskip

\noindent (iii) If $D=D_{\alpha}\kappa_0^{1-\alpha}$ and $\mu <1$ then $\calM_{\rm gam}(\alpha)<1$ and it is a monotonically increasing function of $\alpha$. That is, the more subdiffusive the particle motion is (smaller $\alpha$), the greater the reduction in the MFPT. On the other hand, if $\mu >1$ then $\calM_{\rm gam}(\alpha)>1$ and it is a monotonically decreasing function of $\alpha$.  In this case, more subdiffusive motion results in a greater increase in the MFPT. Both scenarios are illustrated in Figs. \ref{fig3} and \ref{fig4} for the 3D and 2D cases, respectively.

\section{Conclusion}

In this paper we continued our development of an encounter-based approach to combining subdiffusion with non-Markovian mechanisms of absorption. In the case of adsorption at a surface, see Fig. \ref{fig1}(a), the relevant object of interest is the local time propagator, which was analyzed in our companion paper \cite{PCBI}. On the other hand, as shown here, the relevant object for a partially absorbing trap, see Fig. \ref{fig1}(b),  is the occupation time propagator. Both propagators evolve according to fractional diffusion equations, each of which can be derived by taking the appropriate continuum limit of a corresponding Feynman-Kac equation for a heavy-tailed CTRW. One of the major findings of our combined work is that the effects of subdiffusion and non-Markovian absorption on FPT problems are intermingled. For example, in the case of surface adsorption, the FPT density exhibits power-law behavior at large times. If the local time threshold density $\psi(\ell)$ is itself heavy-tailed, then both subdiffusion and adsorption contribute to the power law \cite{PCBI}. In this paper we showed that the effects of these two processes on the MFPT for absorption by a trap, assuming it exists, are also intermingled in the sense that  their contributions are non-separable. 

One issue that we do not address is the numerical implementation of encounter-based reaction-subdiffusion models. A number of numerical methods have been developed to solve time-fractional diffusion equations with Dirichlet, Neumann and Robin boundary conditions \cite{Yuste06,Lin07,Murio08,Zhang14,Herrera21}. They typically involve some form of finite difference scheme, possibly combined with spectral methods. In order to extend these schemes to non-Markovian models of adsorption/absorption, it is necessary to include a numerical algorithm for evaluating the particle-absorbent contact time, that is, the local or occupation time. We have recently developed an efficient algorithm for evaluating the boundary local time that is based on a so-called Skorokhod integral representation of the latter, which we applied to a snapping out Brownian motion model of diffusion through semi-permeable interfaces \cite{Schumm23}. It should be possible to incorporate this algorithm into a numerical scheme for simulated encounter-based reaction-subdiffusion models.
Finally, it would also be interesting in future work to consider other examples of anomalous diffusion such as super-diffusive processes in which the size of jumps of a CTRW are taken to be L\'evy flights \cite{Metzler04}.


\end{document}